\documentstyle[12pt]{article}
\catcode`\@=11

\@addtoreset{equation}{section}
\def\theequation{\arabic{equation}}

\def\@normalsize{\@setsize\normalsize{15pt}\xiipt\@xiipt
\abovedisplayskip 14pt plus3pt minus3pt%
\belowdisplayskip \abovedisplayskip
\abovedisplayshortskip  \z@ plus3pt%
\belowdisplayshortskip  7pt plus3.5pt minus0pt}
\def\small{\@setsize\small{13.6pt}\xipt\@xipt
\abovedisplayskip 13pt plus3pt minus3pt%
\belowdisplayskip \abovedisplayskip
\abovedisplayshortskip  \z@ plus3pt%
\belowdisplayshortskip  7pt plus3.5pt minus0pt
\def\@listi{\parsep 4.5pt plus 2pt minus 1pt
            \itemsep \parsep
            \topsep 9pt plus 3pt minus 3pt}}

\def\underline#1{\relax\ifmmode\@@underline#1\else
        $\@@underline{\hbox{#1}}$\relax\fi}
\@twosidetrue
\relax

\catcode`@=12

\catcode`\@=11

\def\section{\@startsection{section}{1}{\z@}{3.5ex plus 1ex minus
   .2ex}{2.3ex plus .2ex}{\large\bf}}

\def\ps@headings{\def\@oddfoot{}\def\@evenfoot{}
\def\@oddhead{\hbox{}\hfill
        \makebox[.5\textwidth]{\raggedright\ignorespaces --\thepage{}--
        \hfill }}
\def\@evenhead{\@oddhead}
\def\subsectionmark##1{\markboth{##1}{}}
}

\ps@headings

\catcode`\@=12

\relax

\def\figcap{\section*{Figure Captions\markboth
        {FIGURECAPTIONS}{FIGURECAPTIONS}}\list
        {Fig. \arabic{enumi}:\hfill}{\settowidth\labelwidth{Fig. 999:}
        \leftmargin\labelwidth
        \advance\leftmargin\labelsep\usecounter{enumi}}}
 \relax
\def\tablecap{\section*{Table Captions\markboth
        {TABLECAPTIONS}{TABLECAPTIONS}}\list
        {Table \arabic{enumi}:\hfill}{\settowidth\labelwidth{Table 999:}
        \leftmargin\labelwidth
        \advance\leftmargin\labelsep\usecounter{enumi}}}
 \relax
\def\reflist{\section*{References\markboth
        {REFLIST}{REFLIST}}\list
        {[\arabic{enumi}]\hfill}{\settowidth\labelwidth{[999]}
        \leftmargin\labelwidth
        \advance\leftmargin\labelsep\usecounter{enumi}}}
 \relax

\catcode`\@=11

\def\marginnote#1{}
\newcount\hour
\newcount\minute
\newtoks\amorpm
\hour=\time\divide\hour by60
\minute=\time{\multiply\hour by60 \global\advance\minute by-
\hour}
\edef\standardtime{{\ifnum\hour<12 \global\amorpm={am}%
    \else\global\amorpm={pm}\advance\hour by-12 \fi
    \ifnum\hour=0 \hour=12 \fi
    \number\hour:\ifnum\minute<100\fi\number\minute\the\amorpm}}
\edef\militarytime{\number\hour:\ifnum\minute<100\fi\number\minute}
\def\draftlabel#1{{\@bsphack\if@filesw {\let\thepage\relax
  \xdef\@gtempa{\write\@auxout{\string
    \newlabel{#1}{{\@currentlabel}{\thepage}}}}}\@gtempa
    \if@nobreak \ifvmode\nobreak\fi\fi\fi\@esphack}
     \gdef\@eqnlabel{#1}}
\def\@eqnlabel{}
\def\@vacuum{}
\def\draftmarginnote#1{\marginpar{\raggedright\scriptsize\tt#1}}
\def\draft{\oddsidemargin -.5truein
        \def\@oddfoot{\sl preliminary draft \hfil
        \rm\thepage\hfil\sl\today\quad\militarytime}
        \let\@evenfoot\@oddfoot \overfullrule 3pt
        \let\label=\draftlabel
        \let\marginnote=\draftmarginnote
   
\def\@eqnnum{(\theequation)\rlap{\kern\marginparsep\tt\@eqnlabel}%
\global\let\@eqnlabel\@vacuum}  }
\def\preprint{\twocolumn\sloppy\flushbottom\parindent 1em
        \leftmargini 2em\leftmarginv .5em\leftmarginvi .5em
        \oddsidemargin -.5in    \evensidemargin -.5in
        \columnsep 15mm \footheight 0pt
        \textwidth 250mmin      \topmargin  -.4in
        \headheight 12pt \topskip .4in
        \textheight 175mm
        \footskip 0pt
        
\def\@oddhead{\thepage\hfil\addtocounter{page}{1}\thepage}
        \let\@evenhead\@oddhead \def\@oddfoot{} \def\@evenfoot{} 
}
\def\titlepage{\@restonecolfalse\if@twocolumn\@restonecoltrue\onecolumn
     \else \newpage \fi \thispagestyle{empty}\c@page\z@
        \def\thefootnote{\fnsymbol{footnote}} }
\def\endtitlepage{\if@restonecol\twocolumn \else  \fi
        \def\thefootnote{\arabic{footnote}}
        \setcounter{footnote}{0}}  
\catcode`@=12
\relax

\def\ps@headings{\def\@oddfoot{}\def\@evenfoot{}
\def\@oddhead{\hbox{}\hfill
        \makebox[.5\textwidth]{\raggedright\ignorespaces --\thepage{}--
        \hfill }}
\def\@evenhead{\@oddhead}
\def\subsectionmark##1{\markboth{##1}{}}
}

\ps@headings

\relax

\def\firstpage#1#2#3#4#5#6{
\begin{document}
\begin{titlepage}
\nopagebreak
\title{\begin{flushright}
        \vspace*{-1.8in}
        {\normalsize CPTH--S465.0996}\\[-9mm]
        {\normalsize IEM-FT-143/96}\\[-9mm]
        {\normalsize hep-th/9609209}\\[4mm]
\end{flushright}
\vfill
{#3}}
\author{\large #4 \\[1.0cm] #5}
\maketitle
\vskip -7mm     
\nopagebreak 
\begin{abstract}
{\noindent #6}
\end{abstract}
\vfill
\begin{flushleft}
\rule{16.1cm}{0.2mm}\\[-3mm]
$^{\star}${\small Research supported in part by\vspace{-4mm}
IN2P3-CICYT under contract Pth 96-3,
in part by the EEC contracts \vspace{-4mm}
CHRX-CT92-0004 and CHRX-CT93-0340, and in part by 
CICYT contract AEN95-0195.}\\[-3mm] 
$^{\dagger}${\small Laboratoire Propre du CNRS UPR A.0014.}\\
September 1996
\end{flushleft}
\thispagestyle{empty}
\end{titlepage}}

\def\simlt{\stackrel{<}{{}_\sim}}
\def\simgt{\stackrel{>}{{}_\sim}}
\newcommand{\dal}{\raisebox{0.085cm}
{\fbox{\rule{0cm}{0.07cm}\,}}}
\newcommand{\dt}{\partial_{\langle T\rangle}}
\newcommand{\dtbar}{\partial_{\langle\bar{T}\rangle}}
\newcommand{\al}{\alpha^{\prime}}
\newcommand{\mst}{M_{\scriptscriptstyle \!S}}
\newcommand{\mpl}{M_{\scriptscriptstyle \!P}}
\newcommand{\dv}{\int{\rm d}^4x\sqrt{g}}
\newcommand{\lv}{\left\langle}
\newcommand{\rv}{\right\rangle}
\newcommand{\ph}{\varphi}
\newcommand{\abar}{\bar{a}}
\newcommand{\sbar}{\,\bar{\! S}}
\newcommand{\xbar}{\,\bar{\! X}}
\newcommand{\fbar}{\,\bar{\! F}}
\newcommand{\zbar}{\bar{z}}
\newcommand{\dbar}{\,\bar{\!\partial}}
\newcommand{\tbar}{\bar{T}}
\newcommand{\taubar}{\bar{\tau}}
\newcommand{\ubar}{\bar{U}}
\newcommand{\ybar}{\bar{Y}}
\newcommand{\phb}{\bar{\varphi}}
\newcommand{\cm}{Commun.\ Math.\ Phys.~}
\newcommand{\prl}{Phys.\ Rev.\ Lett.~}
\newcommand{\pr}{Phys.\ Rev.\ D~}
\newcommand{\pl}{Phys.\ Lett.\ B~}
\newcommand{\ibar}{\bar{\imath}}
\newcommand{\jbar}{\bar{\jmath}}
\newcommand{\np}{Nucl.\ Phys.\ B~}
\newcommand{\F}{{\cal F}}
\renewcommand{\L}{{\cal L}}
\newcommand{\A}{{\cal A}}
\newcommand{\e}{{\rm e}}
\newcommand{\be}{\begin{equation}}
\newcommand{\en}{\end{equation}}
\newcommand{\gsi}{\,\raisebox{-0.13cm}{$\stackrel{\textstyle
>}{\textstyle\sim}$}\,}
\newcommand{\lsi}{\,\raisebox{-0.13cm}{$\stackrel{\textstyle
<}{\textstyle\sim}$}\,}
\date{}
\firstpage{3118}{IC/95/34}
{\large\bf Large radii and string unification$^{\star}$} 
{I. Antoniadis$^{\,a}$ and M. Quir\'os$^{\,b}$}
{\normalsize\sl
$^a$Centre de Physique Th\'eorique, Ecole Polytechnique,$^\dagger$
{}F-91128 Palaiseau, France\\[-3mm]
\normalsize\sl
$^b$Instituto de Estructura de la Materia, CSIC, Serrano 123, 28006 Madrid,
Spain.}
{We study strong coupling effects in four-dimensional heterotic string models
where supersymmetry is spontaneously broken with large internal dimensions,
consistently with perturbative unification of gauge couplings. These effects
give rise to thresholds associated to the dual theories: type I superstring or
M-theory. In the case of one large dimension, we find that these thresholds
appear close to the field-theoretical unification scale $\sim 10^{16}$ GeV,
offering an appealing scenario for unification of gravitational and gauge
interactions. We also identify the inverse size of the eleventh dimension of
M-theory with the energy at which four-fermion effective operators become
important.}

Large internal dimensions in string theories have been studied in connection with
perturbative breaking of supersymmetry \cite{ablt}--\cite{b}. Their inverse size
is proportional to the scale of supersymmetry breaking which is expected to be of
the order of the electroweak scale. The existence of such large dimensions is
consistent with perturbative unification in a class of four-dimensional models
which include some orbifold compactifications of the heterotic superstring
\cite{a}. Present experimental limits have been obtained from an analysis of
effective four-fermion operators, yielding $R^{-1}\simgt 200$ GeV or
1 TeV in the case of one or two large dimensions, respectively \cite{ab}. The
main experimental signature of these models is the direct production of
Kaluza-Klein excitations for gauge bosons which can be detected at future
colliders \cite{abq}.

The presence of large internal dimensions implies that the ten-dimensional
heterotic string is strongly coupled \cite{kds}. In spite of this, in the class of
models mentioned above, the radiative corrections to the four-dimensional
couplings remain small \cite{a}. This happens for instance when the corresponding
Kaluza-Klein excitations are organized in multiplets of $N=4$ supersymmetry
(possibly spontaneously broken to $N=2$ \cite{kkpr}) leading to cancellations
among particles of different spins. However, the fact that the ten-dimensional
coupling is strong raises the question of possible large corrections to other
quantities of the four-dimensional effective field theory, such as
non-renormalizable operators \cite{ab,ckm}.  Although this problem is difficult to
handle using perturbative methods, it can be studied using recent results
on string dualities. 

There is a growing evidence that the strongly coupled heterotic string in ten
dimensions is equivalent to the weakly coupled type I superstring \cite{pw} or to
the eleven-dimensional M-theory \cite{hw}. The corresponding duality relations
imply the existence of different thresholds associated to these dual theories
where the effective theory changes regime. These thresholds may also appear as
energy scales at which non-renormalizable operators become important \cite{ckm}.
An analysis of type I superstring and M-theory thresholds in models with six
large internal dimensions reveals that these thresholds appear much below the
compactification scale $R^{-1}$, implying for the latter a lower bound $\sim
4\times 10^7$ GeV \cite{ckm}. 

In this letter we study strong coupling effects in the class of models of
refs.~\cite{a}--\cite{ab} with anisotropic compactification space and where
supersymmetry breaking is induced by the large internal dimension(s). We find
that the threshold of dual theories appear now much above the compactification
scale. Moreover, in the case of one large dimension at the TeV range, these
thresholds are close to the experimentally inferred unification scale
$\sim 10^{16}$ GeV, while the inverse size of the eleventh dimension of
M-theory is at an intermediate scale $\sim 10^{13}$ GeV. This offers an
alternative economical scenario for unification of gravitational and gauge
interactions in the context of open strings or M-theory\footnote{For a different
approach see ref.~\cite{w}.}. In fact both the unification and supersymmetry
breaking scales, along with the electroweak one, can in principle be determined
by a single dynamical calculation in the low energy theory \cite{amq}.

For type I strings, we establish the precise relation between the open string
scale and the unification mass in the low energy theory. We also analyze the role
that non-renormalizable operators play concerning the different thresholds. We
find that, while the dimension eight operators $F^4_{\mu\nu}$ lead to the
threshold of type I superstrings, the dimension six four-fermion operators
reproduce the threshold of the eleventh dimension of M-theory, providing
additional evidence for heterotic -- M-theory duality.

\begin{center}
\section*{Type I superstring threshold}
\end{center}

In the heterotic string, the ten-dimensional string coupling $\lambda_H$ and the
string scale $M_H\equiv{\alpha'}_H^{-1/2}$ are expressed in terms of
four-dimensional parameters as:
\be
\lambda_H=2(\alpha_G V)^{1/2}M_H^3 \qquad\qquad 
M_H=\left({\alpha_G\over 8}\right)^{1/2}M_P\ ,
\label{lHMH}
\en
where $(2\pi)^6 V$ is the volume of the six-dimensional internal manifold,
$\alpha_G$ is the gauge coupling at the unification scale and
$M_P=G_N^{-1/2}$ is the Planck mass. Using the experimental values
$M_P=1.2\times 10^{19}$ GeV and $\alpha_G\sim 1/25$ (assuming minimal
supersymmetric unification), one finds $M_H\sim 10^{18}$ GeV while $\lambda_H$
grows to huge values as the internal volume gets large.

In ten dimensions, the heterotic $SO(32)$ and type I strings are related by
duality as \cite{pw}:
\be
\lambda_I={1\over\lambda_H} \qquad\qquad
M_I= M_H\lambda_H^{-1/2}\ .
\label{HI}
\en
This implies that in the limit of large volume, the dual type I superstring is
weakly coupled ($\lambda_I\ll 1$) and its scale $M_I\ll M_H$. In terms of four
dimensional parameters,
\be
M_I=\left({\sqrt{2}\over\alpha_G M_P}\right)^{1/2}V^{-1/4}\ .
\label{MI}
\en
When the internal manifold is large and isotropic, $V=R^6$ and the type I
threshold $M_I\sim (\alpha_G M_P)^{-1/2}R^{-3/2}$ is much below the
compactification scale $R^{-1}$ \cite{ckm}. In this case, it might be possible
to lower the open string scale down to the TeV region, which would be
the threshold of a genuine four-dimensional string \cite{l}. All phenomenology
should then be reexamined.
The open string scale $M_I$ appears also as the threshold at which the dimension
eight effective operators $F_{\mu\nu}^4$ become important \cite{ckm}. On the
heterotic side these operators receive contributions at the one loop level, while
on the type I side they arise at the tree level \cite{t}.

Equation (\ref{MI}) shows that the situation is reversed for anisotropic
internal manifolds with less than four large dimensions. The open string scale
becomes now larger than the compactification scale. In particular, for the class
of models where supersymmetry breaking is tied to the size of just one large
dimension, $V\sim R$ and
\be
M_I= \left( {\sqrt{2}\over\alpha_G}\right)^{1/2}
r^{-5/4}R^{-1/4}M_P^{3/4}\ ,
\label{MI1D}
\en
where $r$ is the size of the five ``small" internal radii in units of $M_P$. For
$R^{-1}=1$ TeV and $r={\cal O}(1)$, one obtains $M_I\sim 7\times 10^{15}$ GeV
which is very close to the gauge coupling unification scale. In this way, when
going up in energies, the physical picture is the following. Between the TeV and
the unification scale the effective theory can be studied using perturbation
theory in the heterotic string. It behaves as five dimensional but with
peculiarities related to the orbifold character of the compactification and the
mechanism of supersymmetry breaking \cite{a}--\cite{ab}. In particular, chiral
states (quarks and leptons) do not have Kaluza-Klein excitations, while the
couplings run with the energy as in four dimensions. At the unification scale, the
theory becomes a genuine type I string weakly coupled.

In order to make precise the relation of the open string scale $M_I$ with the
unification mass, one has to take into account the string threshold corrections
to gauge couplings which can be computed on the type I side for any particular
model. A direct one loop computation in the string theory gives \cite{bf}:
\be
{4\pi\over\alpha_i}={4\pi\over\alpha_G}+\int{dt\over t}{\cal B}_i(t)\ ,
\label{thr}
\en
where $i$ labels the gauge group factor and the integrand ${\cal B}_i$
depends on the specific model. In the ultraviolet limit $t\to 0$,
${\cal B}_i\sim 1/t +{\cal O}(e^{-1/t})$ and the integral appears to have a
quadratic divergence. This is the only short-distance divergence which can be
present in string theory and it is associated to tadpoles of massless
particles. The tadpole cancellation implies a particular
regularization of the different open string diagrams which makes the
result ultraviolet finite \cite{bf}. It consists to cutoff the contributions from
the annulus at $t=1/\Lambda^2$ and from the M\"obius strip at $t=1/4\Lambda^2$.
This leads to the prescription:
\be
{\displaystyle \int_0{dt\over t}{\cal B}_i\equiv \lim_{\Lambda\to\infty}\left\{
{4\over 3}\int_{1/\Lambda^2}{\cal B}_i-
{1\over 3}\int_{1/4\Lambda^2}{\cal B}_i\right\} }\ .
\label{UV}
\en

The integral (\ref{thr}) still has a logarithmic infrared divergence,
since as $t\to\infty$, ${\cal B}_i$ goes to a constant $b_i$. This is a physical
divergence which reproduces the correct low-energy running of the gauge couplings
$\alpha_i$ with beta function coefficients $b_i$. It can
be regularized by introducing an infrared cutoff at
$t=1/\alpha'_I\mu^2$. To compare the string expression with the field
theoretical couplings in a particular renormalization scheme, we have
to add in the r.h.s. of eq.~(\ref{thr}) an appropriate constant term \cite{k}.
In the $\overline{\rm DR}$ renormalization scheme, or equivalently in
the Pauli-Villars (PV) scheme \cite{drpv}, the constant is:
\be
{\displaystyle \lim_{\mu\to 0} \left\{ b_i\ln {\Lambda^2_{\rm PV}\over\mu^2}
-b_i\int_0^{1/\alpha'_I\mu^2}{dt\over t}\left( 
1-e^{ -\pi\alpha'_I\Lambda^2_{\rm PV}t}\right)\right\}
=-b_i\ln (\pi e^\gamma ) }\ ,
\label{const}
\en
where $\gamma\simeq 0.6$ is the Euler's constant. By adding the
constant (\ref{const}) in eq.~(\ref{thr}) one finds:
\be
{4\pi\over\alpha_i}={4\pi\over\alpha_G}+ b_i\ln{M_I^2\over\mu^2} +
\Delta^I_i \ ,
\label{coupl}
\en
where the type I string threshold corrections $\Delta^I_i$ in the $\overline{\rm
DR}$ scheme are given by:
\be
{\displaystyle \Delta^I_i=\lim_{\varepsilon\to 0} \left\{
\int_0^{1/\varepsilon} {dt\over t}{\cal B}_i(t) + b_i\ln\varepsilon \right\}
-b_i\ln (\pi e^\gamma ) }\ .
\label{deltaI}
\en

Notice that the threshold corrections in the heterotic string have a similar
expression as an integral over the complex modular parameter $\tau$ of the
world-sheet torus. The main difference is that the ultraviolet divergence is now
regularized by the restriction to the fundamental domain $\Gamma$ of the modular
group. The limit $\varepsilon\to 0$ in eq.~(\ref{deltaI}) can then be taken easily
by subtracting and adding $b_i$ to the integrand ${\cal B}_i$:
\begin{eqnarray}
\Delta^H_i &=& {\displaystyle\int_\Gamma{d^2\tau\over{\rm Im}\tau}({\cal
B}_i(\tau,\taubar)-b_i) +b_i\lim_{\varepsilon\to 0} \left\{\int_\Gamma^{
1/\varepsilon} {d^2\tau\over{\rm Im}\tau} + \ln\varepsilon \right\}-b_i\ln
(\pi e^\gamma ) }\nonumber\\
&=& \int_\Gamma{d^2\tau\over{\rm Im}\tau}({\cal B}_i(\tau,\taubar)-b_i)
+b_i\ln{2e^{1-\gamma}\over\pi\sqrt{27}}\ .
\label{deltaH}
\end{eqnarray}
The last constant can be absorbed in the definition of the string unification
scale (see eq.~(\ref{coupl})), while the remaining integral defines the threshold
corrections \cite{k}. 

Such a procedure cannot be adopted in the open string case,
since it gives rise to artificial logarithmic ultraviolet divergences which
cannot be regularized by the prescription (\ref{UV}). However, eq.~(\ref{deltaI})
provides a well defined way for computing threshold corrections in type I
string models. For the class of models we consider in this work, threshold
effects depend mainly on the size of the ``small" dimensions $r$ and can provide
model dependent corrections to the unification scale $M_I$.

\vspace{1cm}

\begin{center}
\section*{M-theory threshold}
\end{center}

The strong coupling limit of the heterotic $E_8\times E'_8$ superstring in ten
dimensions is believed to be described by the eleven-dimensional M-theory
compactified on the semi-circle $S^1/Z_2$ of radius $\rho$ \cite{hw}. The
relations between the eleven- and ten-dimensional parameters are:
\be
M_{11}=M_H \left( {\sqrt{2}\over\lambda_H}\right)^{1/3}\qquad\qquad
\rho^{-1}={1\over\sqrt{2}\lambda_H}M_H\ ,
\label{HM}
\en
where we have defined the eleven-dimensional scale $M_{11}=2\pi
(4\pi\kappa^2)^{-1/9}$ \cite{ckm}. When the ten-dimensional heterotic coupling is
large ($\lambda_H\gg 1$), the radius of the semi-circle is large and M-theory is
weakly coupled on the world-volume. 

Using eq.~(\ref{lHMH}), one can express $M_{11}$ and $\rho$ in
terms of the four-dimensional parameters:
\be
M_{11}=(2\alpha_G V)^{-1/6}\qquad\qquad
\rho^{-1}=\left( {2\over\alpha_G}\right)^{3/2}M_P^{-2}V^{-1/2}\ .
\label{M11}
\en
It follows that the M-theory threshold $M_{11}$ is always above (or of the order
of) the compactification scale $R^{-1}$. Furthermore, in the case of one large
dimension, $V\sim R$ and the scale of the eleventh dimension $\rho^{-1}$ is also
larger than $R^{-1}$:
\be
M_{11}=(2\alpha_G R)^{-1/6}r^{-5/6} M_P^{5/6}\qquad\qquad
\rho^{-1}=\left( {2\over\alpha_G}\right)^{3/2}r^{-5/2}R^{-1/2}M_P^{1/2}\ .
\label{M111D}
\en
For $R^{-1}=1$ TeV and $r={\cal O}(1)$, one obtains $M_{11}\sim 3\times 10^{16}$
GeV which essentially coincides with the gauge coupling unification
scale\footnote{A precise connection between $M_{11}$ and the field theoretical
unification mass needs a genuine calculation in the M-theory which is not
available at present. Here, we assume that the unification mass is reasonably
approximated by $M_{11}$, as it was the case for the type I threshold.}.
Moreover, the eleventh dimension threshold is at the intermediate scale
$\rho^{-1}\sim 4\times 10^{13}$ GeV. Thus, in the region between
the TeV and the intermediate scale, the effective five dimensional theory has a
perturbative heterotic string description (as in the case of type I strings
below $M_I$). Above the intermediate scale, strong coupling effects are relevant
and the eleventh dimension of M-theory opens up. Finally, at the unification
scale, gravity becomes important through M-theory interactions.

The situation is reversed if there are more than two large dimensions in the
internal volume $V$. The threshold of the eleventh dimension $\rho^{-1}$ is now
below the compactification scale $R^{-1}$ \cite{bd}. In this case, as going up in
energies, the theory would first become five-dimensional at the scale
$\rho^{-1}$, while the other large dimensions will open up at a higher scale
$R^{-1}< M_{11}$. Moreover, in all these regions the theory does not have a
perturbative string description. For the case of two large dimensions $R^{-1}\sim
10^{-2}\rho^{-1}$, while for the case of six $R^{-1}\sim M_{11}$.

One may ask the question whether the scale of the eleventh dimension
$\rho^{-1}$ can appear on the heterotic side as a threshold at which some
non-renormalizable effective operators become important, in a similar way as the
open string threshold $M_I$ was determined from an analysis of the dimension eight
operators $F_{\mu\nu}^4$. In the following we will argue that the dimension six
four-fermion operators are the relevant ones. 

Let us consider indeed the effective interaction of four chiral
fermions corresponding to twisted states in orbifold models with $2\leq d\leq 6$
large internal dimensions of common size $R$. At energies below $R^{-1}$ the
(tree-level) result can be obtained directly in the effective field theory by
summing over all Kaluza-Klein excitations exchanged between the two fermion lines
\cite{ab}. In the case where all fermions arise at the same fixed point of the
orbifold, the coupling of two twisted states with one excited (untwisted) mode,
labeled by the $d$-dimensional vector ${\vec n}$, is \cite{orb}:
\be
{\displaystyle g_{\vec n}=g\; \delta^{-{\vec n}^2\alpha'_H/2R^2}}\ ,
\label{gn}
\en
where $g$ is the four-dimensional string coupling and the value of the
constant $\delta\geq 1$ depends on the orbifold. The strength $\xi^2$ of the
corresponding effective operator can be written as:
\be
{\displaystyle \xi^2=\alpha_G R^2\sum_{\{\vec{n}\}\neq 0}
{\delta^{ -{\vec n}^2\alpha'_H/R^2}\over{\vec n}^2}
\; \sim\; \alpha_G R^d{\alpha'}_H^{ (2-d)/2}  }\ ,
\label{xi}
\en
in the large $R$ limit. In terms of four-dimensional parameters, using
eq.~(\ref{lHMH}) and the relation ${\textstyle V\sim R^d {\alpha'}_H^{(6-d)/2}}$,
one finds:
\be
\xi^2\sim \alpha_G^3M_P^4V\ .
\label{xi2}
\en
The scale $\xi^{-1}$ defines the energy threshold at which the dimension six
four-fermion operators become important. Experimental bounds on $\xi^{-1}$ are
obtained by identifying it with the scale of compositeness and they yield
typically $\xi^{-1}\simgt 1$ TeV \cite{ab}. 

Note that the expression (\ref{xi2}) for $\xi^{-1}$ is identical to the scale of
the M-theory eleventh dimension (\ref{M11}). We believe this is not an
accident but a consequence of heterotic -- M-theory duality.
In fact, as we discussed above, on the M-theory side the lowest threshold is
$\rho^{-1}$ and to first approximation the four-fermion operator receives
contributions only from the exchange of the Kaluza-Klein modes associated to this
single extra dimension. A similar computation as in eq.~(\ref{xi}) for $d=1$,
$R=\rho$ and $g_n={\cal O}(1)$, then gives $\xi\sim \rho$. 

To make the above argument, it is crucial that the heterotic string is
compactified on an orbifold with twisted sectors, implying on the M-theory
side that the internal seven-dimensional space is not a product of the orbifold
with the semi-circle and $Z_2$ has a non-trivial action. Otherwise, if the
heterotic string were compactified on a smooth manifold $M_6$ (or on an orbifold
without fixed points), its dual model would be, by an adiabatic argument,
M-theory compactified on $M_6\times S^1/Z_2$ \cite{vw,w}.  In this case, ordinary
matter originated from $E_8\times E'_8$ arises at the two fixed points of the
semi-circle and  has only gravitational interactions with Kaluza-Klein states
associated to the eleventh dimension \cite{ckm}. Therefore, the above
four-fermion operator could not be used to extract information on the scale
$\rho^{-1}$.

Finally, it would be interesting to consider, in the context of M-theory, the
possibility of breaking spontaneously $N=1$ supersymmetry using the radius of the
eleventh dimension by a mechanism analogous to the Scherk-Schwarz compactification
\cite{ss}. Here, there are two possibilities:
\begin{itemize}
\item
In the case where M-theory is compactified on a seven-dimensional space which does
not contain the semi-circle as a product factor, the scale of supersymmetry
breaking in the observable sector ($m_{\rm susy}$) would be generically
proportional to $\rho^{-1}\sim 1$ TeV. From eq.~(\ref{M11}), the $d$ additional
large dimensions would then open up at an intermediate scale varying between
$10^6$ and $10^{13}$ GeV corresponding to $d=3$ and $d=6$,
respectively. 
\item
In the case where the internal manifold of M-theory is $M_6\times S^1/Z_2$,
supersymmetry is broken only in the gravitational sector (at the lowest order)
and will be communicated to the observable world by gravitational interactions,
yielding $m_{\rm susy}\sim\rho^{-2}/M_P$. As a result, the threshold of the
eleventh dimension $\rho^{-1}$ should be at an intermediate scale $\sim 10^{12}$
GeV. Interestingly enough, for $d=6$ the inverse size of the six-dimensional
internal manifold $M_6$ is now of the order of the gauge coupling unification
mass $\sim 10^{16}$ GeV. It is suggestive that this situation could describe
ordinary gaugino condensation in the dual strongly coupled heterotic 
string~\cite{horava}.
\end{itemize}

\end{document}